\documentclass[aps,pr,twocolumn,superscriptaddress,groupedaddress]{revtex4}  
\usepackage{graphicx}  
\usepackage{dcolumn}   
\usepackage{bm}        
\usepackage{amssymb}   
\usepackage{amsmath}
\usepackage{ascmac}
\usepackage{color}
\begin{document}

\title{Measurement-based quantum computation using two-component BECs}
\input 
\author{Genji Fujii\\Department of Nuclear Engineering, Kyoto University, 6158540 Kyoto, Japan}
\date{\today}

\begin{abstract}
In this paper, we propose measurement-based quantum computation (MBQC) using two-component Bose-Einstein condensates (BECs). Graph states are naturally introduced by analogy with the qubit case. An arbitrary state of one logical qubit can be obtained through a three-body measurement. Furthermore, we propose methods for implementing CZ gates on the components of coherent states of BECs in a graph state, as well as CZ gates on logical qubits. A distinctive feature of our approach is that the post-measurement state is shown to depend on the particle number. These results suggest a novel quantum computing process based on particle control.
   
\end{abstract}

\maketitle

\section{introduction}
\indent In the early 1990s, the concept of the quantum computer began to take shape and attention was drawn to the possibility of applying the principles of quantum mechanics to computational theory, exploring new approaches to problems considered intractable for conventional classical computers. The idea of the quantum simulator, introduced by Richard Feynman, was based on the perspective of using quantum mechanics itself as the foundation of computation. Subsequently, the discovery of numerous useful quantum algorithms demonstrated the potential of quantum computers. Concepts such as qubits and quantum gates came to be recognized as realistic computational models, and the theory of quantum computation developped.\\
\indent Bose-Einstein condensation (BEC) is a phenomenon theoretically predicted by Bose and Einstein. They showed that particles with integer spin (bosons) could accumulate in a macroscopically occupied quantum state at ultralow temperatures, leading to the emergence of quantum behavior on a macroscopic scale. However, this phenomenon was experimentally unobservable with the technology available at that time and remained a theoretical prediction for more than half a century. Later, since its realization in cold atomic systems in 1995 [1,2], BEC has attracted considerable attention both experimentally and theoretically[3]. BEC has been applied to quantum technologies such as quantum simulation [4] and quantum metrology [5]. A further theoretical development was the proposal of quantum computation using BEC [6]. The authors of Ref. [6] employed two modes of the hyperfine state to encode information in BECs. By exploiting appropriate time evolution, they demonstrated that entangled states between two BECs could be created [7]. Promising systems for performing quantum computation through BECs encoding include optical lattices [3], atom chips [8], and yttrium iron garnet (YIG) [9]. Each has its own advantages: optical lattices offer excellent controllability, atom chips provide high coherence, and YIG enables BEC generation even at room temperature.\\
\indent On the other hand, a method of quantum computation called measurement-based quantum computation (MBQC) [10,11] has been proposed. At present, quantum computers have two types as well as a degree of freedom in what is used as a quantum-bit. Regarding the types, the circuit model and the measurement-based model are known, with the circuit model being the first to be proposed. MBQC represents a different computational process from the circuit model and has expanded into distinct research directions. MBQC performs quantum computation by preparing a topological state called a graph state and applying unitary operators through measurements. Graph states are useful quantum states for quantum computation, and many studies have been conducted on hypergraph states [12] using generalized controlled-Z gates between two qubits, as well as qudit graph states [13-15] using d-dimensional Hilbert spaces. Moreover, MBQC has generated concepts not found in the circuit model. Examples include matrix product state representations [16-18], correlation space [19,20], valence bond solid states [21], and projected entangled pair states [22-24]. Furthermore, it is noteworthy that graph states have been experimentally realized in optical lattice systems [25], and whether they can also be realized in atom chip and YIG systems represents an intriguing direction for future research.\\
\indent As pointed out earlier, there is also freedom in the choice of quantum-bits. For example, possible choices include qubits, qudits, CV qubits, anyons, coherent states, and the BECs mentioned previously. What we propose here is measurement-based and employs BECs. In this paper, we project the BECs onto a two-dimensional logical qubit and demonstrate that arbitrary single-qubit rotations can be performed within this space. We refer to this process as a three-body measurement. Furthermore, to establish universality, a two-qubit gate is required. We also show that, by construction, an example of a two-qubit gate, namely the CZ gate, can be implemented within the graph state formalism.\\
\indent The structure of this paper is as follows. In Section II, we review the basic mathematical aspects of MBQC and BECs. In Section III, we examine a new framework of MBQC by combining these two concepts. It is shown that arbitrary rotations on a two-dimensional logical qubit can be realized by the three-body measurement, and that universal quantum computation can be achieved in combination with the two-type CZ gates.
\section{preliminaries}
\subsection{MBQC and Graph States}
\ \ \ \ \ \ \ \ \ \ \ \ \ \ \ \ \ \includegraphics[width=50mm]{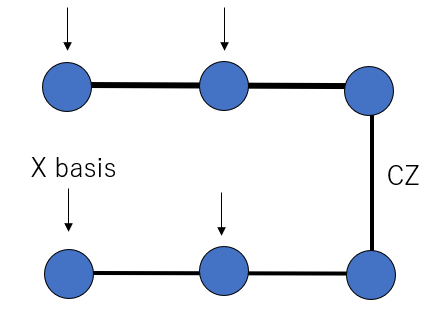}\\
\ \  Fig1: The CZ gate can be implemented beforehand preparation since the measurement and the CZ gate are commuted.\\
\\
In MBQC, the computational resource known as a graph state consists of vertices and the edges connecting them. The general definition of a graph state is given mathematically by assigning the state 
$|+\rangle = \frac{(|1\rangle +|0\rangle )}{\sqrt{2}}$ to each vertex and using CZ gates as the edges. That is, a graph state is defined as 
\begin{equation}
|G\rangle\equiv\left(\prod_{e\in E} CZ_{e}\right)|+\rangle^{\otimes|V|},
\end{equation}
where $CZ_{e}$ affects two vertices belonging to edge $e$, and V is the vertex. Edge addition and deletion can be performed by the CZ gate. \\
\indent Any qubit state, on the Bloch sphere representation, can be expressed using a graph state after measurement. Consider an explicit two-qubit case as an example. The prepared graph states are 
\begin{equation}
|G\rangle=|0\rangle_{1}|+\rangle_{2}+|1\rangle_{1}|-\rangle_{2}.
\end{equation} 
\indent Then qubit 1 was measured in the basis ${|0\rangle_{1} \pm e^{i\theta}|1\rangle_{1}}$. If it is projected onto $|0\rangle+e^{i\theta}|1\rangle$, the remaining qubit is found in a state that has undergone the action of the following unitary gate.
\begin{equation}
\begin{split}
&|+\rangle_{2}+e^{-i\theta}|-\rangle_{2}\\
&=He^{i\theta Z/ 2}|+\rangle_{2},
\end{split}
\end{equation}
where $H$ is the Hadamard gate. Any one-qubit unitary operator can be realized with $H$ and $e^{i\theta Z/ 2}$ . \\
\indent Since the measurement and the CZ gate can be interchanged, by measuring the $X$-basis for the four qubits on the left, the remaining state can be regarded as the state after computation in which the CZ gate remains on the two qubits at the right end (Fig. 1).
\subsection{Two-component BECs}
 Quantum computation using BECs was proposed [6]. BECs quantum computations express computational process by two levels of the hyperfine state via creation operators $a^{\dagger}$ and $b^{\dagger}$. Arbitrary BECs qubit states are
\begin{equation}
|\alpha,\beta\rangle\rangle\equiv\frac{1}{\sqrt{N!}}(\alpha a^{\dagger}+\beta b^{\dagger})^{N}|\rm{vac}\rangle.
\end{equation}
\indent Here $a^{\dagger}$ and $b^{\dagger}$ are creation Bose operators, obeying commutation relations $[a,a^{\dagger}]=[b,b^{\dagger}]=1$.  $\alpha and \beta$ are arbitrary complex numbers satisfying $|\alpha|^{2}+|\beta|^{2}=1$ and $|\rm{vac}\rangle$ in a vacuum state. The boson number $\mathcal{N}=a^{\dagger}a+b^{\dagger}b$ is taken to be the conserved quantity. To manipulate two-component BECs, the Stokes operator is introduced.

\begin{equation}
\begin{split}
S^{x}&=a^{\dagger}b+b^{\dagger}a\\
S^{y}&=-ia^{\dagger}b+ib^{\dagger}a\\
S^{z}&=a^{\dagger}a-b^{\dagger}b.
\end{split}
\end{equation}
\indent Those operators satisfy the commutation relation with same spin-$1/2$ particles, $[S^{i},S^{j}]=2i\epsilon_{ijk}S^{k}$ where $\epsilon_{ijk}$ is the Levi-Civita antisymmetric tensor $i, j, k = x, y, z$. An entangled state generated by $S_{1}^{z}S_{2}^{z}$ interaction is generated by
\begin{equation}
\begin{split}
&e^{-iS^{z}_{1}S^{z}_{2}t}|\frac{1}{\sqrt{2}},\frac{1}{\sqrt{2}}\rangle\rangle_{1}|\frac{1}{\sqrt{2}},\frac{1}{\sqrt{2}}\rangle\rangle_{2}\\
&=\frac{1}{\sqrt{2^{N}}}\sum_{k_{2}=0}^{N}\sqrt{\begin{pmatrix}N\\k_{2}\end{pmatrix}}|\frac{e^{-i(2k_{2}-N)t}}{\sqrt{2}},\frac{e^{i(2k_{2}-N)t}}{\sqrt{2}}\rangle\rangle_{1}|k_{2}\rangle_{2},
\end{split}
\end{equation}
where $k$ is a Fock state defined by 
\begin{equation}
\begin{split}
|k\rangle \equiv\frac{(a^{\dagger})^{k}(b^{\dagger})^{N-k}}{\sqrt{k!(N-k)!}}|vac\rangle.
\end{split}
\end{equation}
Another important operator is a BEC-type CZ gate, which acts on two BECs and whose Hamiltonian is described by
\begin{equation}
H=-S^{z}_{1}S^{z}_{2}+\mathcal{N}_{2}S_{1}^{z}-\mathcal{N}_{1}S_{2}^{z}+\mathcal{N}_{1}\mathcal{N}_{2},
\end{equation}
which affects two BECs and for a time evolution $t=\pi/4N$. This gives
\begin{equation}
\frac{1}{\sqrt{2^{N}}}\sum_{k_{2}=0}^{N}\sqrt{\begin{pmatrix}N\\k_{2}\end{pmatrix}}|\frac{1}{\sqrt{2}},\frac{e^{\frac{-i\pi k_{2}}{N}}}{\sqrt{2}}\rangle\rangle_{1}
|k_{2}\rangle_{2}.
\end{equation}
\section{MBQC with Two-Component BECs}
\subsection{Formalism}
\indent In this chapter, we introduce a representation using BECs within the framework of MBQC. First, we define the graph state by analogy with the qubit version of MBQC. In the MBQC formalism with BECs, we place $|\frac{1}{\sqrt{2}},\frac{1}{\sqrt{2}}\rangle\rangle\ $ instead of $|+\rangle$. A at the vertices of the graph state, and construct the graph state by applying the CZ gate between the BECs located at the vertices. In this paper, we fix the time evolution of the CZ gate to
 $t=\pi / 4$.  Let us define two types of BECs CZ gates. Those that are Hamiltonian are defined by\\
\begin{equation}
H_{1}=-S^{z}_{1}S^{z}_{2}+\mathcal{N}_{2}S_{1}^{z}-\mathcal{N}_{1}S_{2}^{z}+\mathcal{N}_{1}\mathcal{N}_{2}
\end{equation} 
\begin{equation}
H_{2}=S^{z}_{1}S^{z}_{2}+\mathcal{N}_{2}S_{1}^{z}+\mathcal{N}_{1}S_{2}^{z}+\mathcal{N}_{1}\mathcal{N}_{2}.
\end{equation}
  A Hamiltonian $H_{1}$, time evolution acts on two BECs qubits as
\begin{equation}
\begin{split}
&\exp(-iH_{1}t)|\frac{1}{\sqrt{2}},\frac{1}{\sqrt{2}}\rangle\rangle_{1}|\frac{1}{\sqrt{2}},\frac{1}{\sqrt{2}}\rangle\rangle_{2}\\
&=\frac{1}{\sqrt{2^{N}}}\sum_{k_{1}=0}^{N}\sqrt{\begin{pmatrix}N\\k_{1}\end{pmatrix}}|k_{1}\rangle_{1}|\frac{1}{\sqrt{2}},\frac{e^{-i\pi k_{1}}}{\sqrt{2}}\rangle\rangle_{2}.
\end{split}
\end{equation} 
\indent This type of Hamiltonian time evolution is referred to as a right-hand component phase shift CZ gate (r-CZ).
Conversely, a left-hand component phase shift CZ gate (l-CZ) Hamiltonian $H_{2}$ acts between two BEC qubit as
\begin{equation}
\begin{split}
&\exp(-iH_{2}t)|\frac{1}{\sqrt{2}},\frac{1}{\sqrt{2}}\rangle\rangle_{1}|\frac{1}{\sqrt{2}},\frac{1}{\sqrt{2}}\rangle\rangle_{2}\\
&=\frac{1}{\sqrt{2^{N}}}\sum_{k_{2}=0}^{N}\sqrt{\begin{pmatrix}N\\k_{2}\end{pmatrix}}|\frac{e^{-i\pi k_{2}}}{\sqrt{2}},\frac{1}{\sqrt{2}}\rangle\rangle_{1}|k_{2}\rangle_{2}\\
&=\frac{1}{\sqrt{2^{N}}}\sum_{k_{1}=0}^{N}\sqrt{\begin{pmatrix}N\\k_{1}\end{pmatrix}}|k_{1}\rangle_{1}|\frac{e^{-i\pi k_{1}}}{\sqrt{2}},\frac{1}{\sqrt{2}}\rangle\rangle_{2}.
\end{split}
\end{equation}
\indent The above two CZ gates are described by a similar formalism. However, there are different calculation results after the measurement steps. Hamiltonian $H_{1}$ results in any N-dependent logical qubit state. Conversely, $H_{2}$ is independent. \\
\indent In our formalism, we can define general BECs graph states
\begin{equation}
\begin{split}
|G\rangle\rangle&=\left(\prod_{e\in E}\exp(-iH_{a}t)_{e}\right)|\frac{1}{\sqrt{2}},\frac{1}{\sqrt{2}}\rangle\rangle^{\otimes |V|},\\
\end{split}
\end{equation}
where, the CZ gate $CZ_{e}=\exp(-iH_{a}t)_{e}$, $a\in{1,2}$ affects two vertices belonging to an edge $e$. After discussion, binomial coefficient factors were neglected. For an explicit example of two BECs cases, and after summing over $k_{1}$, the BECs graph states are written as
\begin{equation}
\begin{split}
|G\rangle\rangle=|e\rangle_{1}|\frac{1}{\sqrt{2}},\frac{1}{\sqrt{2}}\rangle\rangle_{2}+|o\rangle_{1}|\frac{1}{\sqrt{2}},\frac{-1}{\sqrt{2}}\rangle\rangle_{2}.
\end{split}
\end{equation}
\ \ \ A total even number ket $|e\rangle$ and a total odd number ket $|o\rangle$ is defined as
\begin{equation}
\begin{split}
|e\rangle\equiv \sum_{k=0}^{k\in even}\sqrt{_{N} C _{k}}|k\rangle,\\
|o\rangle\equiv \sum_{k=1}^{k\in odd}\sqrt{_{N} C _{k}}|k \rangle.
\end{split}
\end{equation}
\indent Equation (15) is a simple extension of equation (2). Indeed, one can confirm this by replacing $|0\rangle \rightarrow |e\rangle $, $|1\rangle \rightarrow |o\rangle $, $|+\rangle \rightarrow|\frac{1}{\sqrt{2}},\frac{1}{\sqrt{2}}\rangle\rangle$, and $|-\rangle \rightarrow |\frac{1}{\sqrt{2}},\frac{-1}{\sqrt{2}}\rangle\rangle$
\subsection{The Three-Body Measurement}
\includegraphics[width=80mm]{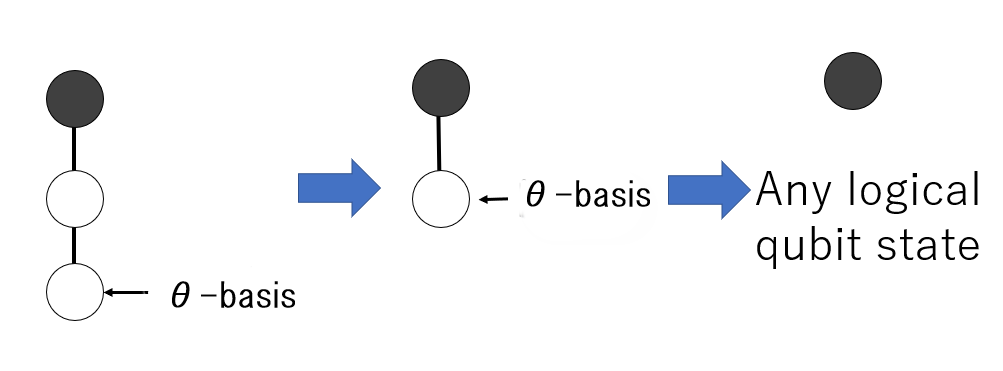}\\
\ \  Fig2: . A black circle denotes BECs before summing over $k$. A white circle denotes executed BECs summing over $k$. Consuming two BECs qubits, one can obtain any logical qubit state.\\
\\
\indent MBQC is an approach that aims to perform quantum computation by making measurements on graph states. There are several types of measurements, e.g., X-measurement, Z-measurement, and measurements in a Hadamard basis with a phase. First, we introduce the BECs version of the Z-measurement. This corresponds to a set of measurements in the Fock basis and is represented by the following set: $\{|0\rangle,.... |k\rangle,.....|N\rangle\}$.\\
\indent We also introduce a measurement corresponding to the X-measurement. This is a basis obtained by applying a unitary to the Fock-basis measurement (the Z-measurement in BECs), and this unitary is defined as
\begin{equation}
\begin{split}
U_{1}=e^{-in^{b}_{3,l}2\pi}e^{-in^{b}_{3,m}\pi}e^{-3iS^{y}\pi/4}.
\end{split}
\end{equation}
\indent The Hadamard gate $(H= e^{-3iS^{y}\pi/4})$ acts on the Fock basis is $(a^{\dagger}+b^{\dagger})^{k}(a^{\dagger}-b^{\dagger})^{N-k}/\sqrt{k!(N-k)!}|{\rm vac}\rangle$. $n^{b}_{3,l}$ and $n^{b}_{3,m}$ act as $n^{b}_{3,l}|l\rangle=(k-l)|l\rangle$, $n^{b}_{3,m}|m\rangle=(N-k-m)|m\rangle$, where  $|l\rangle=(a^{\dagger})^{l}(b^{\dagger})^{k-l}/\sqrt{l!(k-l)!}|{\rm vac}\rangle$ and $|m\rangle=(a^{\dagger})^{m}(b^{\dagger})^{N-k-m}/\sqrt{m!(N-k-m)!}|{\rm vac}\rangle$.\\
\indent The unitary operator $U_{1}$ transforms the Fock basis into the basis of coherent states $|\frac{1}{\sqrt{2}},\frac{1}{\sqrt{2}}\rangle\rangle$. We define this basis as the X-measurement basis, and it is expressed as follows using the total-odd and total-even ket states $|\frac{1}{\sqrt{2}},\frac{1}{\sqrt{2}}\rangle\rangle=|e\rangle+|o\rangle$, if $N$ is even number. \\
\indent We can also introduce the transformation to the basis orthogonal to this basis. It is obtained by transforming the Fock basis using the following unitary operator.
\begin{equation}
\begin{split}
U_{2}=e^{-in^{b}_{3,l}\pi}e^{-in^{b}_{3,m}2\pi}e^{-3iS^{y}\pi/4}.
\end{split}
\end{equation}
This basis constitutes the  $|\frac{1}{\sqrt{2}},\frac{-1}{\sqrt{2}}\rangle\rangle=|e\rangle-|o\rangle$ measurement basis, if $N$ is even number.\\
\indent The measurement basis with a phase inserted into the Hadamard basis can be constructed as
\begin{equation}
\begin{split}
U_{\theta}=e^{-3iS^{y}\pi/4}e^{-in^{b}_{3,l}\theta}e^{-in^{b}_{3,m}\theta}e^{-3iS^{y}\pi/4}.
\end{split}
\end{equation}
\indent We call this measurement-basis as $\theta$-basis, and express as $|\theta\rangle$\\
\indent The Three-body measurement was showed in (Fig 2). Any rotation on two-dimentional logical Bloch sphere can be realized by the Three-body measurement. The Three-body measurement consists of the following eight steps.\\
\begin{screen}\ \ \ \ \ \ \ \ \ \ \ \ \ [The Three-body measurement]
\\
\\ 1. Implement CZ gate between BECs1 and BECs2. \\
2. Implement CZ gate between BECs2 and BECs3.\\ 
3. Sum over BECs3.\\
4. Measure by $\theta$-basis at BECs3.\\ 
5. Sum over BECs2. \\
6. Measure by $\theta$-basis at BECs2.\\ 
7. Implement BEC-type Hadamard gate on BECs1.\\
8. Projection into logical qubit on BECs1.\end{screen}
\\ \\ 
\indent First, the r-CZ gate is implemented between BECs1 and BECs2. Then, the r-CZ gate is implemented between BECs2 and BECs3. The system becomes
\begin{equation}
|G\rangle\rangle=\sum_{k_{2}=0}^{N}\sum_{k_{3}=0}^{N}|\frac{1}{\sqrt{2}},\frac{e^{-i\pi k_{2}}}{\sqrt{2}}\rangle\rangle_{1}|k_{2} , e^{-i\pi k_{3}}\rangle_{2}|k_{3}\rangle_{3},
\end{equation}
where 
\begin{equation}
|k,e^{-i\pi k'}\rangle\equiv\frac{(a^{\dagger})^{k}(b^{\dagger}e^{-i\pi k'})^{N-k}}{\sqrt{k!(N-k)!}}|\rm{vac}\rangle.
\end{equation}
\indent A summation is taken over $k_{3}$ in eq.(20). Thus the graph state as
\begin{equation}
\begin{split}
&\sum_{k_{2}=0}^{N}\big(|\frac{1}{\sqrt{2}},\frac{e^{-i\pi k_{2}}}{\sqrt{2}}\rangle\rangle_{1}|-k_{2}\rangle_{2}|o\rangle_{3}\\
&+|\frac{1}{\sqrt{2}},\frac{e^{-i\pi k_{2}}}{\sqrt{2}}\rangle\rangle_{1}|k_{2}\rangle_{2}|e\rangle_{3}\big).
\end{split}
\end{equation}
\indent The next step is measurement on BECs3 by $\theta$-basis. Here, it is necessary to carefully consider how the $\theta$-measurement acts on the odd-ket and even-ket. First, note that for an $x$-polarized BECs coherent state, the $\theta$-measurement acts as
\begin{equation}
\begin{split}
\langle\theta |\frac{1}{\sqrt{2}},\frac{1}{\sqrt{2}}\rangle\rangle&=(\frac{\cos\theta+i\sin\theta}{\sqrt{2}})^{N}\\
&=\sqrt{2^{N}}e^{i\theta N},
\end{split}
\end{equation}
and
\begin{equation}
\begin{split}
\langle\theta |\frac{1}{\sqrt{2}},\frac{-1}{\sqrt{2}}\rangle\rangle&=(\frac{\cos\theta-i\sin\theta}{\sqrt{2}})^{q_{3}}(\frac{i\sin\theta-\cos\theta}{\sqrt{2}})^{N-q_{3}}\\
&=(-1)^{N-q_{3}}(\frac{\cos\theta-i\sin\theta}{\sqrt{2}})^{N}\\
&=\sqrt{2^{N}}(-1)^{N-q_{3}}e^{-i\theta N},
\end{split}
\end{equation}
where $q_{3}$ is a measurement result. Taking these considerations into account, and noting that the even-ket and odd-ket can be expressed as superposition states of $x$-polarized coherent states, the $\theta$-measurement yields:
\begin{equation}
\begin{split}
\langle \theta |e\rangle&=\langle\theta|\left(|\frac{1}{\sqrt{2}},\frac{1}{\sqrt{2}}\rangle\rangle+|\frac{1}{\sqrt{2}},\frac{-1}{\sqrt{2}}\rangle\rangle\right)\\
&=\sqrt{2^{N}}(e^{i\theta N}+(-1)^{N-q_{3}}e^{-i\theta N}),
\end{split}
\end{equation}
and
\begin{equation}
\begin{split}
\langle \theta |o\rangle&=\langle\theta|\left(|\frac{1}{\sqrt{2}},\frac{1}{\sqrt{2}}\rangle\rangle-|\frac{1}{\sqrt{2}},\frac{-1}{\sqrt{2}}\rangle\rangle\right)\\
&=\sqrt{2^{N}}(e^{i\theta N}-(-1)^{N-q_{3}}e^{-i\theta N}).
\end{split}
\end{equation}
Here, depending on the measurement outcome $q_{3}$, the factor applied to these post-measurement states varies as either cosine or sine. Returning to Eq. (22), suppose a $\theta$-measurement is performed and $N_{3}-q_{3}$ is even number. The state after appropriate renormalization is
\begin{equation}
\begin{split}
&\sum_{k_{2}=0}^{N}(i\sin(\theta N_{3})|\frac{1}{\sqrt{2}},\frac{e^{-i\pi k_{2}}}{\sqrt{2}}\rangle\rangle_{1}|-k_{2}\rangle_{2}\\
&+\cos(\theta N_{3})|\frac{1}{\sqrt{2}},\frac{e^{-i\pi k_{2}}}{\sqrt{2}}\rangle\rangle_{1}|k_{2}\rangle_{2}).
\end{split}
\end{equation}
\indent The same procedure was followed for BECs2, if BECs2 measurement result is $N_{2}-q_{2}$ is even, and the measurement angle is $\phi$. The graph state is projected on
\begin{equation}
\begin{split}
\cos(\phi N_{2})|\frac{1}{\sqrt{2}},\frac{1}{\sqrt{2}}\rangle\rangle_{1}+ie^{-2i\theta N_{3}}\sin(\phi N_{2})|\frac{1}{\sqrt{2}},\frac{-1}{\sqrt{2}}\rangle\rangle_{1}.
\end{split}
\end{equation}
\indent Next, the BEC-type Hadamard gate is implemented on BECs1.
\begin{equation}
\begin{split}
e^{-i3\pi S^{y} / 4}|\frac{1}{\sqrt{2}},\frac{1}{\sqrt{2}}\rangle\rangle &=|1,0\rangle\rangle ,\\
 e^{-i3\pi S^{y} / 4}|\frac{1}{\sqrt{2}},\frac{-1}{\sqrt{2}}\rangle\rangle &=|0,1\rangle\rangle.
\end{split}
\end{equation}
\indent  Finally, the BECs1 (N+1 dimensional Hilbert space) is projected into a logical qubit (2-dimensional Hilbert space), 
\begin{equation}
|\tilde0\rangle \equiv |1,0\rangle\rangle,|\tilde1\rangle\equiv |0,1\rangle\rangle.
\end{equation} 
\indent Thus, any single-qubit state can be implemented after the three-body measurement. An outcome state depends on $N_{2}$ and $N_{3}$. This result implies that the outcome state is determined beforehand in the prepared boson number. After all steps are implemented, the following outcome states are obtained, ignoring the overall phase. outcome state is
\begin{equation}
\big(\cos(\phi N_{2})|\tilde{0}\rangle_{1}\\
+ie^{-2i\theta N_{3}}\sin(\phi N_{2})|\tilde{1}\rangle_{1}\big)\\.
\end{equation}
This state can represent any state on the Bloch sphere specified by $\theta$ and $\phi$.
\subsection{Measurements and Controlled-Z Gates}
\ \ \ \ \ \ \ \ \ \ \ \ \ \ \ \ \ \includegraphics[width=50mm]{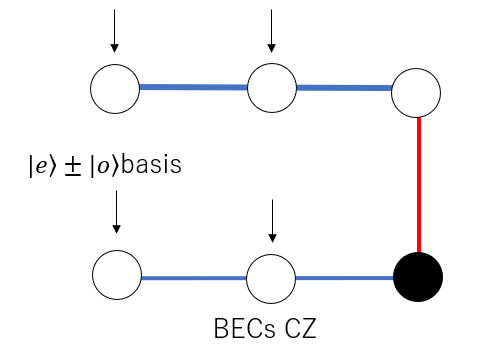}\\
\ \  Fig.3: The black circle denotes BECs before summing over $k$. The white circle denotes executed BECs summing over $k$. A blue line represents the l-CZ, and a red line represents the r-CZ.\\
\indent \indent \indent \indent \indent \indent \includegraphics[width=50mm]{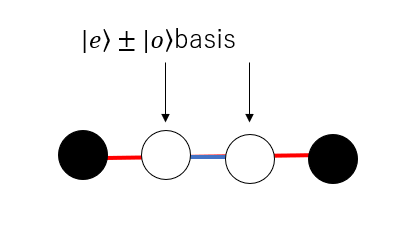}\\
\\
\ \ Fig.4: A logical-type CZ gate can be implemented by measuring the two white circles in the center of the $|e\rangle \pm |o\rangle$ basis. The blue line connects the two white circles and the red line connects the white circle and the black circle.\\
\\
\indent \indent \indent \indent \indent \indent \includegraphics[width=60mm]{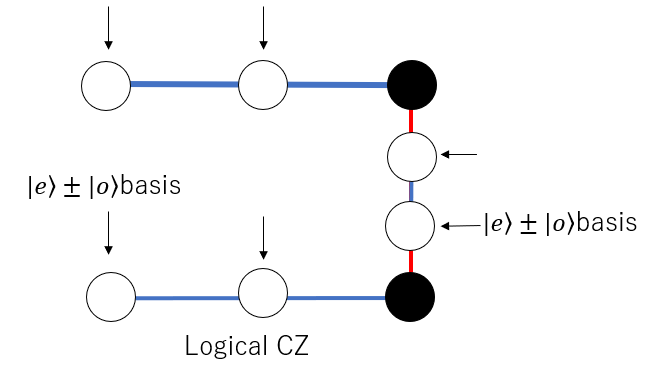}\\
\\
\ \ Fig.5: Implementation of logical-type CZ gate in the BEC graph state.  \\
\\
\indent Using similar procedures for MBQC, the CZ gate can be implemented in our formalism(Fig.3), since the CZ gates can be implemented beforehand preparation. $|e\rangle_{3} \pm |o\rangle_{3}$ basis is projected onto the left four BEC qubits. Note that the two BECs CZ gates (r-CZ or l-CZ) must be carefully chosen. This type CZ gate introduce phase for BECs coherent states. \\
\indent On the other hand, a logical-type CZ gate (Fig4) can also be implemented. First, the four-BEC qubit graph state is
\begin{equation}
\begin{split}
|G\rangle\rangle=&\sum_{k_{2}=0}^{N}\sum_{k_{3}=0}^{N}|\frac{1}{\sqrt{2}},\frac{e^{-i\pi k_{2}}}{\sqrt{2}}\rangle\rangle_{1}| e^{-i\pi k_{3}} , k_{2} \rangle_{2}\\
&\otimes|k_{3}\rangle_{3}|\frac{1}{\sqrt{2}},\frac{e^{-i\pi k_{3}}}{\sqrt{2}}\rangle\rangle_{4},
\end{split}
\end{equation}
\includegraphics[width=90mm]{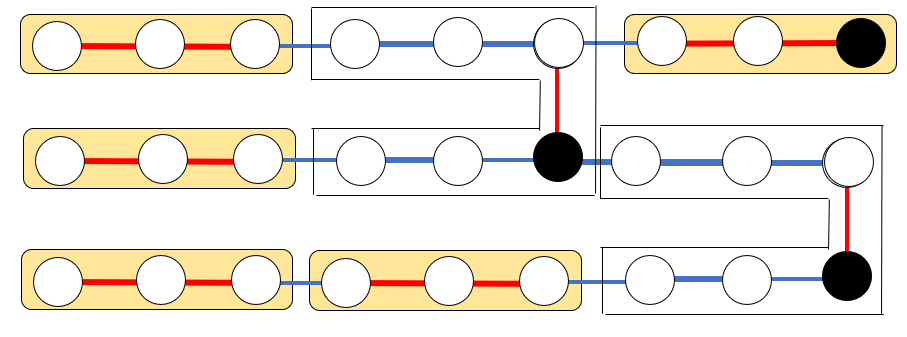}\\
\\
\ \  Fig.6: An example of the BECs graph states. The black circle denotes BECs before summing over k. The white circle denotes executed BECs summing over k. The red part represents the three-body measurement part, and the white part represents the BECs-type CZ gate part.
where 
\begin{equation}
|e^{-i\pi k'} , k \rangle\equiv\frac{(a^{\dagger}e^{-i\pi k'})^{k}(b^{\dagger})^{N-k}}{\sqrt{k!(N-k)!}}|\rm{vac}\rangle.
\end{equation}
\indent Then, after taking the sum of $k_{2}$ and $k_{3}$, the two BECs were measured in the middle of the $|e\rangle \pm |o\rangle$ basis. The graph state is
\begin{equation}
\begin{split}
|G\rangle\rangle&=|\tilde0\rangle|\tilde0\rangle+(-1)^{s_{1}}|\tilde0\rangle|\tilde1\rangle+\\
&(-1)^{s_{2}}|\tilde1\rangle|\tilde0\rangle-(-1)^{s_{1}}(-1)^{s_{2}}|\tilde1\rangle|\tilde1\rangle,
\end{split}
\end{equation}
\indent where $s_{1}$ and $s_{2}$ are measurement results. It is the logical-type CZ gate when projected into a logical qubit if $s_{1}=s_{2}=0$. This type CZ gate introduce phase for logical qubit states. The logical-type CZ gate can be implemented in the BECs graph state structure, shown in Figure 5.\\
\indent A general graph state in the MBQC using BECs, which connected the three-body measurement part and the CZ gate part, is shown in Figure 6.
\section{conclusions}
In this paper, we constructed MBQC using BECs and the BECs graph states using the CZ gates with appropriate time evolution. As a result, the maximally entangled state between logical qubits and the arbitrary logical qubit state was obtained by the three-body measurement. The general graph state was constructed by combining the three-body measurement and the CZ gate part. \\
\indent Finally, it is noted that the key to the realization of the proposed theory in a laboratory will depend on the ability to control particle dissipation of the system. Recent experiments on optical lattices are attempting to control dissipation in the cold atomic system such as inelastic collisions and light scattering of particles [26,27] and atom number fluctuations[28,29]. These experiments have the potential to serve as important milestones for the realization of this study.
\input 


\begin{thebibliography}{99}
\bibitem{1}
M. H. Anderson et al. Science. \textbf{269}, 198 (1995)
\bibitem{2}
K. B. Davis et al. Phys. Rev. Lett. \textbf{75}, 3969 (1995)
 \bibitem{3}
\bibitem{4}
I. Buluta, F. Nori, Science. \textbf{326}, 108 (2009)
\bibitem{5}
A. Sorensen et al. Nature. \textbf{409}, 63 (2000)
P. Treutlein et al. Fortschr. Phys. \textbf{54}, 702 (2006)
\bibitem{6}
T. Byrnes, K. Wen, Y. Yamamoto, Phys. Rev. A \textbf{85}, 040306 (2012)
\bibitem{7}
A. N. Pyrkov and T. Byrnes, New J. Phys. \textbf{15}, 093019 (2013)
\bibitem{8}
R. Folman et al. Adv. in At. Mol. Opt. Physics. \textbf{48}, 263 (2002)
\bibitem{9}
A. A. Serga et al. J. of. Phys. D. \textbf{43}, 26 (2010)
 \bibitem{10}
R. Raussendorf, H. J. Briegel, Phys. Rev. Lett \textbf{86}, 5188 (2001)
  \bibitem{11}
R. Raussendorf, D. E. Browne, H. J. Briegel, Phys. Rev. A \textbf{68}, 022312 (2003)
\bibitem{12}
M. Rossi et al. New J. Phys. \textbf{15}, 11 (2013)
\bibitem{13}
A. Keet et al. Phys. Rev. A. \textbf{82}, 6 (2010)
\bibitem{14}
S. Y. Looi et al. Phys. Rev. A. \textbf{78}, 4 (2008)
\bibitem{15}
W. Tang et al. Phys. Rev. Lett \textbf{110}, 10 (2013)
\bibitem{16}
 U.Schollwoeck, Annals of Physics \textbf{326}, 96 (2011)
\bibitem{17}
D. Sauerwein, A. Molnar, J. I. Cirac and B. Kraus, Phys. Rev. Lett. \textbf{123}, 170504 (2019)
\bibitem{18}
D. P. Garcia, F. Verstraete, M. M. Wolf and J. I. Cirac, Quantum. Inf. Compt. \textbf{7}, 401 (2007)
\bibitem{19}
D. Gross, J. Eisert, Phys. Rev. Lett. \textbf{98}, 220503 (2007)
\bibitem{20}
T. Morimae, Phys. Rev. A \textbf{83}, 042337 (2011)
\bibitem{21}
 F. Verstraete, J. I. Cirac, Phys. Rev. Lett. \textbf{70}, 060302 (2004)
\bibitem{22}
F. Verstraete, V.Murg, and J.I.Cirac, Adv. Phys, \textbf{57}, 2, 143 (2008)
\bibitem{23}
T.Nishino, Y. Hieida, K. Okunishi, N. Maeshima, Y. Akutsu and A.Gendiar, Prog. Theor. Phys, \textbf{105}, 409 (2001)
\bibitem{24}
 N.Maeshima, Y. Hieida, Y. Akutsu, T. Nishino and K. Okunishi, Phys. Rev. E \textbf{64}, 016705 (2001)
\bibitem{25}
K. Inaba et al. Phys. Rev. Lett \textbf{112}, 11 (2014)
 \bibitem{26}
T.Tomita et al. Sci. Adv. \textbf{3}, e1701513 (2017)
\bibitem{27}
R. Bouganne et al. Nat. Phys. \textbf{16}, 21 (2020) 
\bibitem{28}
Y. Shalibo et al. Phys. Rev. Lett . \textbf{110}, 100404 (2013)
\bibitem{29}
D. Xu et al. arXiv preprint arXiv:2211.06644 (2022) 

\end{thebibliography}
\end{document}